\documentclass[a4paper,11pt]{article}
\pdfoutput=1 

\usepackage{jcappub} 

\usepackage[T1]{fontenc} 

\title{\boldmath Multi-angle calculation of the matter-neutrino resonance near an accretion disk}

\author{Shashank Shalgar}
\affiliation{Theoretical Division, Los Alamos National Laboratory, Los Alamos, NM 87545, USA}

\emailAdd{shashank@lanl.gov}

\abstract{
We perform a numerical study of the matter-neutrino resonance in a multi-angle calculation in the vicinity of an accretion disk. We assume thermally distributed neutrino and anti-neutrino fields emitted by two-dimensional disk that is homogeneous and isotropic; the electrons are assumed to be at constant density. We compare the the result of this computation to that obtained using single-angle approximation. We investigate the robustness of matter-neutrino resonance in environment surrounding accretion disks by progressively relaxing the single angle approximation. We find that the multi-angle results in the present simplified model do not support a robust resonance mechanism as suggested by the single angle treatment. We also discuss the context under which matter-neutrino resonance may be important in future studies.

}

\begin{document}
\maketitle
\flushbottom

\section{Introduction}
\label{sec:intro}
In the early universe and extreme astrophysical environments like the interior of core-collapse supernovae and the medium surroundings accretion disks, the neutrino fluxes can be large and neutrino-neutrino forward scattering cannot be neglected while calculating neutrino flavor evolution. Neutrino-neutrino forward scattering leads to non-linear equations of motion for flavor evolution and the resulting phenomenology is poorly understood as a consequence. In simplified core-collapse supernova models it was shown that due to the significant contribution of neutrino-neutrino forward scattering, the neutrinos can oscillate with the same frequency, in a synchronized fashion, irrespective of energy. This phenomenon is generally referred to as `collective neutrino oscillations'~\cite{Pantaleone:1994ns,
Duan:2005cp,Duan:2006an,Duan:2006jv,Duan:2007mv,Duan:2008za,
Raffelt:2007xt,EstebanPretel:2007bz,EstebanPretel:2008ni,
Raffelt:2008hr, Dasgupta:2010ae}. The phenomenon of collective neutrino oscillations has also been extensive studied in the early universe~\cite{Harvey:1981cu,Foot:1995qk,Shi:1996ic,Casas:1997gx,
MarchRussell:1999ig,Kawasaki:2002hq,Yamaguchi:2002vw,Shaposhnikov:2008pf,
Gu:2010dg,Johns:2016enc} and in the medium surrounding accretion disks~\cite{Malkus:2014iqa,Zhu:2016mwa,Wu:2015fga,Frensel:2016fge}.

Numerical calculations of collective neutrino oscillations is computationally very challenging even in the case of very simple geometry like `neutrino bulb model' used in the context of core-collapse supernovae~\cite{Duan:2006an,Duan:2006jv,Duan:2007mv,Duan:2008za}. Moreover, some assumptions made in the neutrino bulb model like the imposition of perfect spherical symmetry may not be realistic. Another assumption that is sometimes made in the context of neutrino bulb model is called the `single angle approximation'. In the single-angle approximation, the dependence of neutrino flavor on the emission angle is completely ignored, by either averaging over emission angles to find the self-interaction Hamiltonian, or a single neutrino emission angle is used to represent all the neutrino medium. In absence of single-angle approximation, neutrinos with different emission angles have to be evolved independently which is computationally very expensive. Such a neutrino flavor evolution is generally referred to as `multi-angle calculation'. If the multi-angle calculation does not develop a significant emission angle dependence then the single-angle and multi-angle calculations give very similar neutrino flavor spectra in the neutrino bulb model. 

Apart from core-collapse supernovae collective neutrino oscillations are also important in the vicinity of accretion disks in neutron star mergers. Neutrino flavor oscillations near neutron star mergers have gained a lot of attention in the literature. Neutron star mergers are a viable site for r-process nucleosynthesis and neutrino flavor oscillations can in principle affect the r-process nucleosynthesis rate. R-process nucleosynthesis requires a neutron rich environment and neutrino oscillations can change the electron neutrino to anti-electron neutrino ratio which can in turn affect neutron to proton ratio via $W$-boson exchange. 

In the accretion disk surrounding neutron star mergers, the anti-electron neutrino flux is larger than the electron neutrino flux due to the large number of neutron decays. This is opposite of what we expect in the interior of a core-collapse supernova. The neutrino self-interacting potential thus has opposite sign around neutron star mergers compared to that in the interior of core-collapse supernovae. Around neutron star mergers the self-interaction potential (which will be defined later in the paper), can cancel the matter potential arising due to the neutrino forward scattering off electrons. Due to the non-linear nature of neutrino flavor evolution if the self-interaction potential cancels the matters potential the system is in the state of maximal mixing and remains that way for a significant time. However, all the papers in the literature on this topic assume that single-angle approximation~\cite{Malkus:2014iqa,Zhu:2016mwa,Wu:2015fga,Frensel:2016fge}. In this paper we discuss the matter-neutrino resonance in the context of multi-angle calculation. 

In Sec.~\ref{mnr} we briefly discuss the phenomenon of matter-neutrino resonance using single-angle approximation, and in Sec.~\ref{mmnr} we relax the the single-angle approximation and discuss the system qualitatively. In Sec.~\ref{results} we perform multi-angle numerical simulations to justify conclusions reached in Sec.~\ref{mmnr}. Finally we discuss the future implications of our calculations and conclude in Sec.~\ref{conclusion}.

\section{Matter neutrino resonance}
\label{mnr}
In this section we use a simplified model of accretion disk to briefly recapitulate the phenomenon and matter-neutrino resonance in single-angle approximation. We consider an infinite homogeneous plane to represent the accretion disk. The neutrinos are emitted at an angle of $\vartheta_{0} = 45^{\circ}$ with respect to normal. We label the direction normal to the infinite plane by $z$. In the case of an infinite disk the neutrino flux of number density should be independent of $z$, however in order to artificially include the effect of finiteness of the size of the disk we assume that the neutrino number density falls off as $1/z^{3}$. 

We use two Wigner transformed density matrices (corresponding to neutrino and anti-neutrinos) to describe the flavor content of neutrinos. In two flavor approximation, which we use throughout this paper, the density matrices are $2 \times 2$ matrices, with diagonal elements that encode the fluxes of a given flavor apart from a normalization factor that is dependent on the overall number density of neutrinos.

The evolution of the neutrino flavor along the direction of propagation, $l$, is given by Heisenberg equations,
\begin{eqnarray}
i\frac{\partial \rho(E)}{\partial l} = [H,\rho(E)] \nonumber \\
i\frac{\partial \bar{\rho}(E)}{\partial l} = [\bar{H},\bar{\rho}(E)],
\label{eom:l}
\end{eqnarray}
where, $\rho(E)$ and $\bar{\rho}(E)$ represent the density matrix for neutrinos and density matrix for anti-neutrinos, respectively. The corresponding Hamiltonians that govern the evolution of the density matrices are given by $H$ and $\bar{H}$. Their form will be discussed later in this section. It will be convenient to express the equations of motion in terms of the $z$ instead of $l$. This can be done by simply dividing the right hand side of eqs.~\eqref{eom:l} by $\cos\vartheta_{0}$,
\begin{eqnarray}
i\frac{\partial \rho(E)}{\partial z} = \frac{1}{\cos\vartheta_{0}}[H,\rho(E)] \nonumber \\
i\frac{\partial \bar{\rho}(E)}{\partial z} = \frac{1}{\cos\vartheta_{0}}[\bar{H},\bar{\rho}(E)].
\label{eom:z}
\end{eqnarray}

In the single-angle approximation, the Hamiltonian including the self-interaction potential is given by,
\begin{eqnarray}
H &=& H_{\textrm{vac}}+H_{\textrm{mat}}+H_{\textrm{self}}\\
\bar{H} &=& -H_{\textrm{vac}}+H_{\textrm{mat}}+H_{\textrm{self}}\\ 
H_{\textrm{vac}} &=& \frac{1}{2}
\begin{pmatrix}
-\omega\cos 2 \theta_{\textrm{V}} & \omega \sin 2 \theta_{\textrm{V}}\\
\omega \sin 2 \theta_{\textrm{V}} & \omega \cos 2 \theta_{\textrm{V}}
\end{pmatrix}\\
H_{\textrm{mat}} &=& 
\begin{pmatrix}
V_{\lambda} & 0\\
0 & 0
\end{pmatrix}\\
H^{\textrm{sa}}_{\textrm{self}} &=& \mu(z) \int dE \left(\rho(E) - \bar{\rho}(E) \right) (1-\cos^{2}\vartheta_{0}).
\label{ham}
\end{eqnarray}
Here, $\mu(z)$ is the strength of the self-interaction potential and is proportional to the neutrino number density. We put an artificial $\mu \sim 1/z^{3}$ dependence on neutrino self-interaction potential to study the matter-neutrino potential as mentioned earlier. $\omega=\frac{\Delta m^{2}}{2E}$, $\theta_{\textrm{V}}$ and $V_{\lambda}$ are the vacuum oscillation frequency, vacuum mixing angle and matter potential respectively. The matter potential is a result of charge-current forward scattering of electron type neutrinos off electrons and is equal to $\sqrt{2}G_{F}n_{e}$, where $G_{F}$ is the Fermi constant and $n_{e}$ is the number density of electrons. 
The superscript `sa' in eq.~\eqref{ham} is used to denote its limited validity to single-angle approximation.

As opposed to the matter Hamiltonian, $H_{\textrm{mat}}$ the self-interaction Hamiltonian is non-linear as it depends on the density matrices of the neutrinos. Also unlike the matter Hamiltonian the self-interaction Hamiltonian in general is dependent on the relative direction between the neutrino under consideration and the neutrino medium. In the single-angle approximation the angular dependence of the Hamiltonian only plays a role of scaling the effective neutrino number density by a constant factor.
In eq.~\eqref{ham} we can see that for $\vartheta_{0}=45^{\circ}$ which we use in our numerical calculations, this scaling of effective neutrino number density, ($1-\cos^{2}\vartheta_{0}$), is of order one.

The self-interaction Hamiltonian can have large negative diagonal values for neutrino spectra emitted by accretion disks, due to the excess of anti-electron neutrinos over electron neutrinos. It is possible to define a potential corresponding to self-interaction similar to that for matter Hamiltonian. It should be noted that terms in the Hamiltonian that are proportional to unit matrix are not physical as they only add an overall phase to all flavors. It is possible to define a potential corresponding to the self-interaction Hamiltonian, consisting of only the diagonal elements, but independent of the freedom to add terms proportional to identity,
\begin{eqnarray}
V_{\textrm{self}} = H^{ee}_{\textrm{self}} - H^{xx}_{\textrm{self}},
\label{Vself}
\end{eqnarray}
were, $H^{ee}_{\textrm{self}}$ and $H^{xx}_{\textrm{self}}$ are the 11 and 22 components of the self-interaction Hamiltonian. 

If the negative self-interaction potential $V_{\textrm{self}}$, has the same magnitude as the matter potential, the effective total Hamiltonian would have vanishing diagonal components. 
For $V_{\textrm{self}} \approx V_{\textrm{mat}} \gg \omega$, it is easy to see that the condition for near maximal mixing, $|H^{ex}| \gg |H^{ee}-H^{xx}|$, is satisfied. The phenomenon of matter-neutrino resonance is the locking of the system in this state of maximal mixing. The neutrino flavor spectra evolve in a manner that this lock is maintained. As the magnitude of self-interaction decreases with $z$, at some point the locking between the self-interaction potential and the matter potential can no longer be sustained and this is clearly seen in the numerical simulations carried out in the papers on the topic~\cite{Malkus:2014iqa,Zhu:2016mwa,Wu:2015fga,Frensel:2016fge} and the orange line in Fig.~\ref{mainfig}. 

\section{Multi-angle matter-neutrino resonance}
\label{mmnr}

The interesting phenomenology arising out of non-linear neutrino flavor evolution has not been tested under relaxation of the single angle approximation. In this section we discuss the formalism we use to relax the single-angle approximation and study the consequences on matter-neutrino resonance. 

In order to study the impact of single-angle approximation we relax the assumption by replacing the single emission angle $\vartheta_{0}$ by a range
\begin{eqnarray}
\vartheta_{0} \rightarrow [\vartheta_{\textrm{min}},\vartheta_{\textrm{max}}] \quad \textrm{and }\Delta \vartheta \equiv \vartheta_{\textrm{max}}-\vartheta_{\textrm{min}}.
\end{eqnarray}
We assume that the flux is uniformly distributed over the range of emission angle and that the total flux is kept the same while relaxing from single-angle approximation to multi-angle calculation.

This change leads to some qualitative differences in the flavor evolution of the system, which we will discuss in this section. In the next section we present numerical results that justify the qualitative claims made in this section.

The change to the equations of motion and self-interaction Hamiltonian that can be seen in the equations below,
\begin{eqnarray}
i\frac{\partial \rho(E)}{\partial z} &=& \frac{1}{\cos\vartheta_{0}}\left[H,\rho(E)\right]
\rightarrow 
i\frac{\partial \rho(E,\cos\vartheta)}{\partial z} = \frac{1}{\cos\vartheta}\left[H,\rho(E,\cos\vartheta)\right] \\
i\frac{\partial \bar{\rho}(E)}{\partial z} &=& \frac{1}{\cos\vartheta_{0}}\left[\bar{H},\bar{\rho}(E)\right]
\rightarrow 
i\frac{\partial \bar{\rho}(E,\cos\vartheta)}{\partial z} = \frac{1}{\cos\vartheta}\left[\bar{H},\bar{\rho}(E,\cos\vartheta)\right] \\
H^{\textrm{sa}}_{\textrm{self}} 
&\rightarrow & 
H^{\textrm{ma}}_{\textrm{self}} = \mu(z) \int dE d(\cos\vartheta^{\prime}) \left(\rho(E,\cos\vartheta^{\prime}) - \bar{\rho}(E,\cos\vartheta^{\prime}) \right) (1-\cos\vartheta^{\prime}\cos\vartheta).
\label{satoma}
\end{eqnarray}
The superscript `ma' in eq.~\eqref{satoma} is used to denoted the multi-angle self-interaction Hamiltonian. 
The form of the Hamiltonian corresponding to the vacuum and matter term remains unchanged.

In the case of multi-angle calculation there are two major differences with respect to single-angle calculation.
The self-interaction potential in the case of multi-angle calculations is a function of emission angle.
In multi-angle equations of motion it is easy to see that the diagonal components of the matter potential and self-interaction potential can never cancel each other for all values to $\vartheta$, but such a cancellation is possible for a particular value to $\vartheta$. In addition the angle dependent factor of $\frac{1}{\cos\vartheta}$ also spoils the matter-neutrino resonance. It is possible to stare at eqs.~\eqref{satoma} and notice that matter-neutrino resonance is not sustainable in the case of multi-angle calculation.

It should however be noted that in multi-angle calculations, the flavor oscillations can be dependent on the emission angle $\vartheta$, and since this angle dependence is related to neutrino mass hierarchy the opening angle, $\Delta \vartheta$, needed to completely get rid of matter-neutrino resonance can be different for normal and inverted hierarchy. That is what we see in the numerical calculations presented in the next section.

\section{Numerical results}
\label{results}
\begin{figure}
\includegraphics[width=0.48\textwidth]{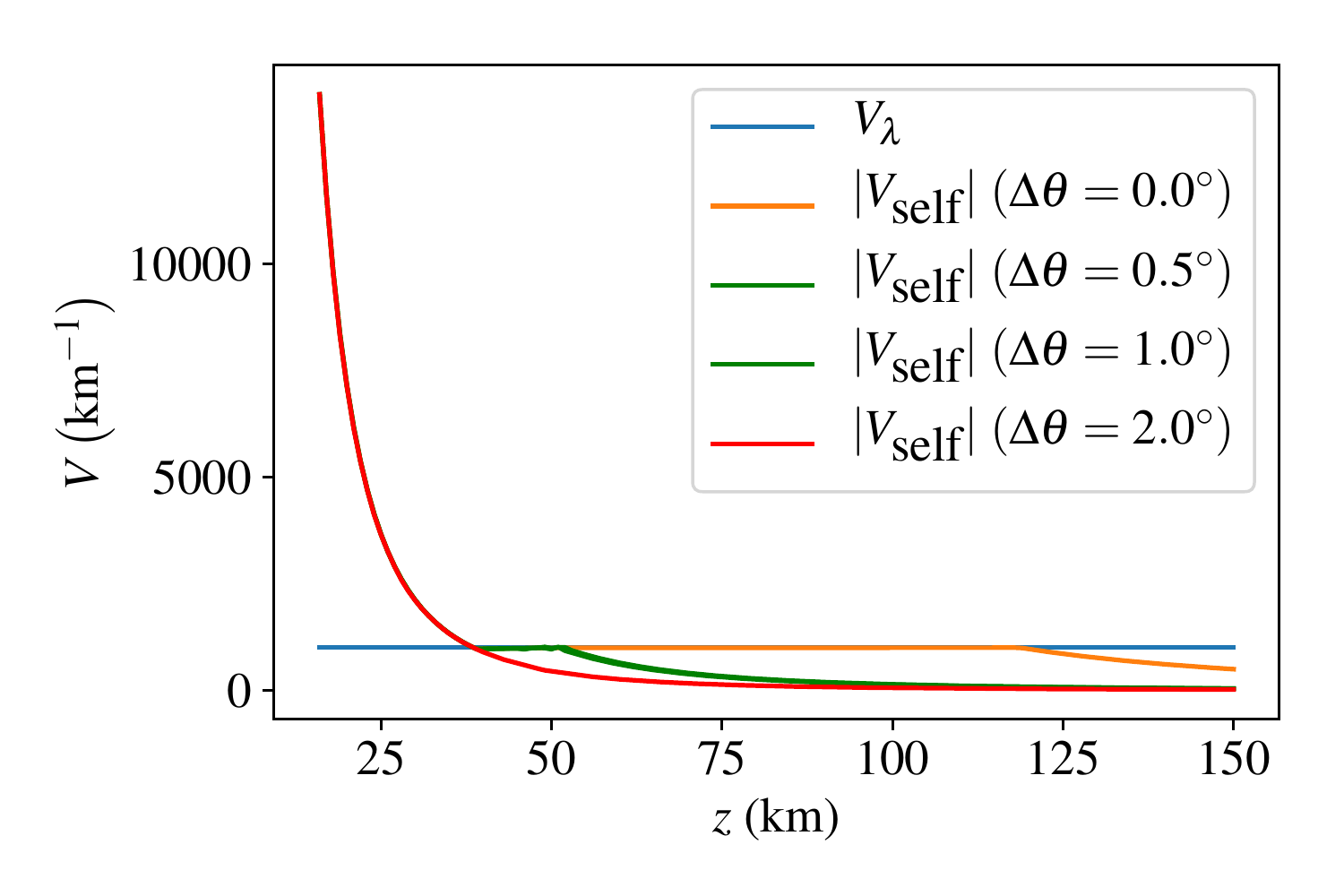}
\includegraphics[width=0.48\textwidth]{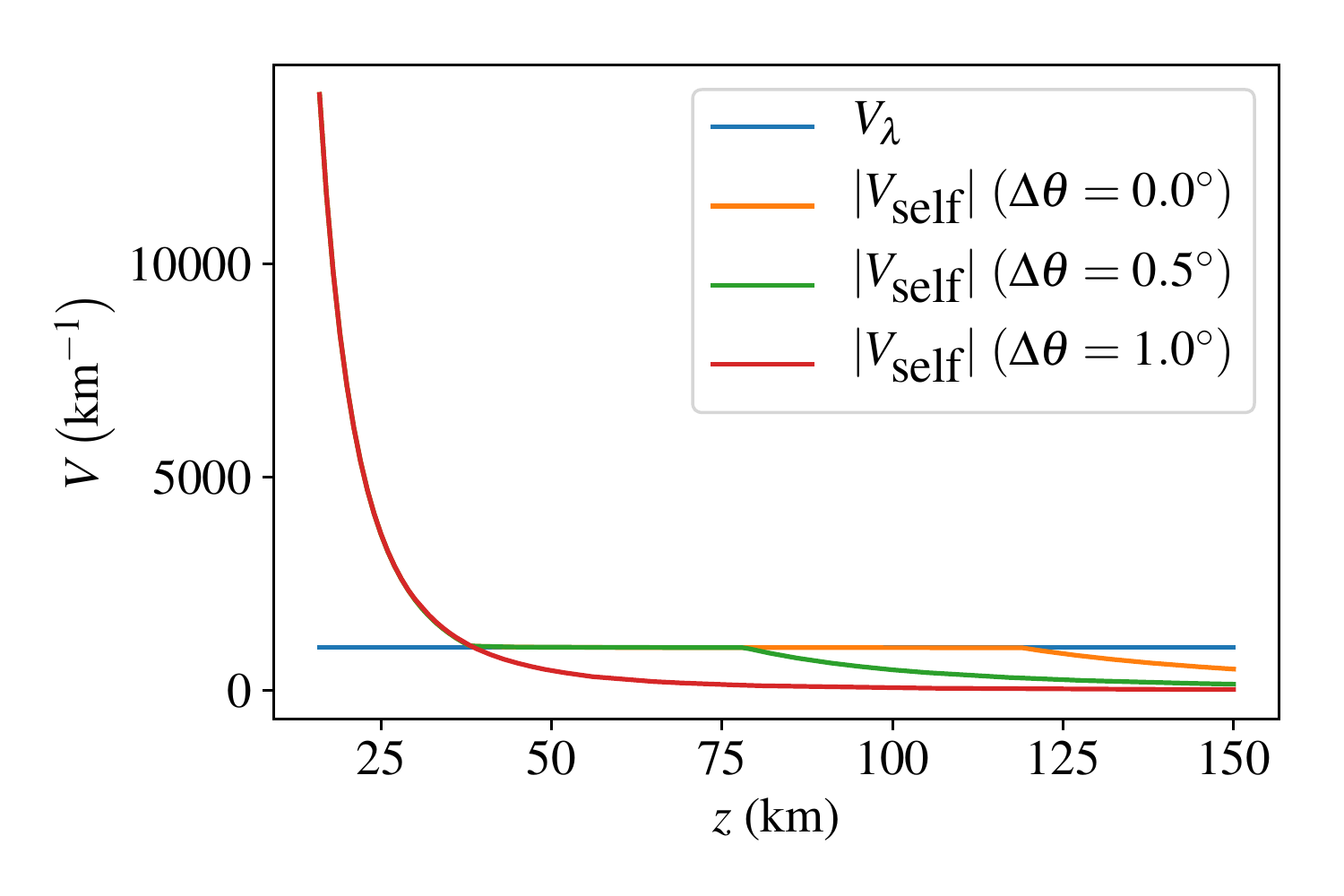}
\caption{Evolution of matter and self-interaction potential for normal(left) and inverted(right) hierarchy for various opening angles. In the case of normal mass hierarchy the evolution of the self-interaction potential is identical for $\Delta\vartheta=0.5^{\circ}$ and $\Delta\vartheta=1.0^{\circ}$ as far as the eyes can see. There is a matter-neutrino resonance which is independent of the opening angle for a small range of opening angles.}
\label{mainfig}
\end{figure}
In this section we verify the qualitative analysis of matter-neutrino resonance using numerical results. For the sake of simplicity we restrict ourselves to an infinite disk that is perfectly homogeneous and isotropic. We assume that the electron number density remains constant in the region where neutrino oscillations happen. Although this is not a realistic density profile it does not affect the assertions made in this paper in any qualitative way.

We assume that the neutrinos emitted by the accretion disk have a thermal distribution, 
\begin{eqnarray}
f_{i}(E) = \frac{E^{2}}{1+\exp\left(\frac{E}{T_{i}}+\eta_{i}\right)},
\end{eqnarray}
where, $E$ is the energy and the subscript $i$ is the flavor index which can be either $\nu_{e}$, $\bar{\nu}_{e}$
or $\nu_{x}$. In medium surrounding the accretion disk the spectrum of $\bar{\nu}_{x}$ is the same as $\nu_{x}$.
The luminosities, temperatures and chemical potentials for various flavors used in our numerical calculation are given in the table below

\begin{tabular}{|l|l|l|l|}
\hline
 		& $\nu_{e}$	& $\bar{\nu}_{e}$	& $\nu_{x}$	\cr
\hline
\hline
Luminosity, $L_{i}$ (ergs/sec) 	& $3.0\times 10^{52}$	& $4.5\times 10^{52}$	& $9\times 10^{51}$	\cr
\hline
Temperature, $T_{i}$ (MeV)	& 2.65		& 3.85			& 4.35		\cr
\hline
chemical potential, $\eta_{i}$ (no units) & 3		& 3 			& 3 		\cr		
\hline
\end{tabular}
\\
Notice that the luminosity of $\bar{\nu}_{e}$ is much larger than that of ${\nu}_{e}$, which makes the total self-interaction potential negative. We hold the matter potential constant in the region of interest at $\lambda=1000$ km$^{-1}$ ($\approx 3.16\times 10^{-22}$ ergs). Fig.~\ref{mainfig} has been obtained using $|\Delta m^{2}|=2.5 \times 10^{-3}$ eV$^{2}$ and $\theta_{V}=0.01$ rad.

For the purpose of plotting, we define multi-angle self-interaction potential evaluated at the midpoint of the range used for multi-angle calculations,
\begin{eqnarray}
V_{\textrm{self}} = \left(H^{ee}_{\textrm{self}}(\cos\vartheta)-H^{xx}_{\textrm{self}}(\cos\vartheta)\right)\Big|_{\vartheta
=\frac{\vartheta_{\textrm{min}}+\vartheta_{\textrm{max}}}{2}}
\end{eqnarray}
The matter potential, $V_{\lambda}$, which is a constant in our calculation is also plotted for reference.

We can see that as we gradually relax the single angle approximation the matter-neutrino resonance becomes less prominent. In the case of inverted mass hierarchy the matter-neutrino resonance completely disappears by opening angle of $1^{\circ}$. By that we mean that the evolution of $V_{\textrm{self}}$ is identical with or without flavor oscillations. In the case of normal hierarchy, however, there is a resonance that is unchanged for moderately small opening angles. The evolution of self-interaction potential is nearly identical for opening angles of $0.5^{\circ}$ and $1^{\circ}$ in Fig.~\ref{mainfig}. However, for the parameters we are working with this region of resonance also disappears by opening angle of $2^{\circ}$. The difference in the behavior of normal and inverted hierarchy can be attributed to emission angle dependent flavor evolution.

\begin{figure}
\includegraphics[width=0.49\textwidth]{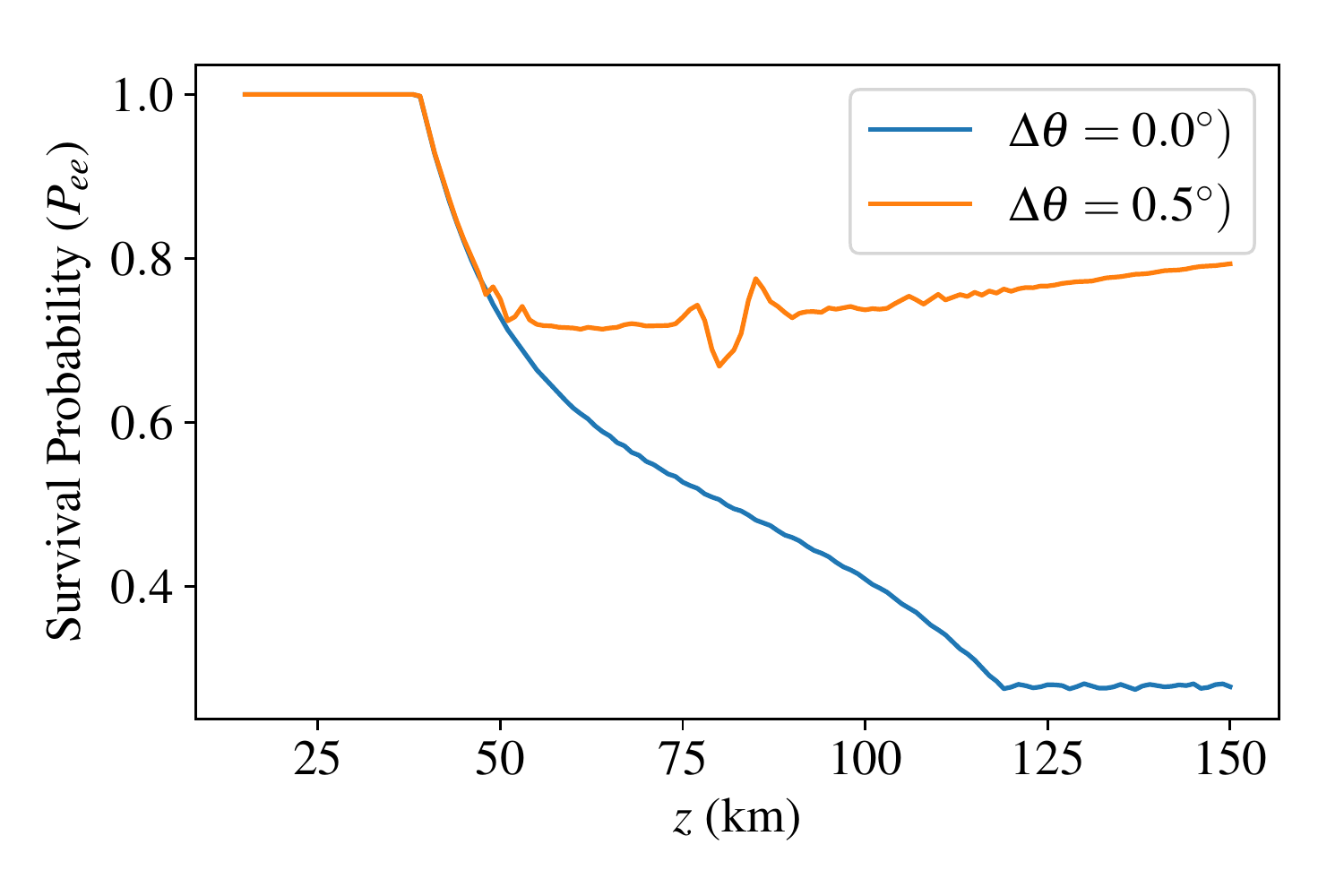}
\includegraphics[width=0.49\textwidth]{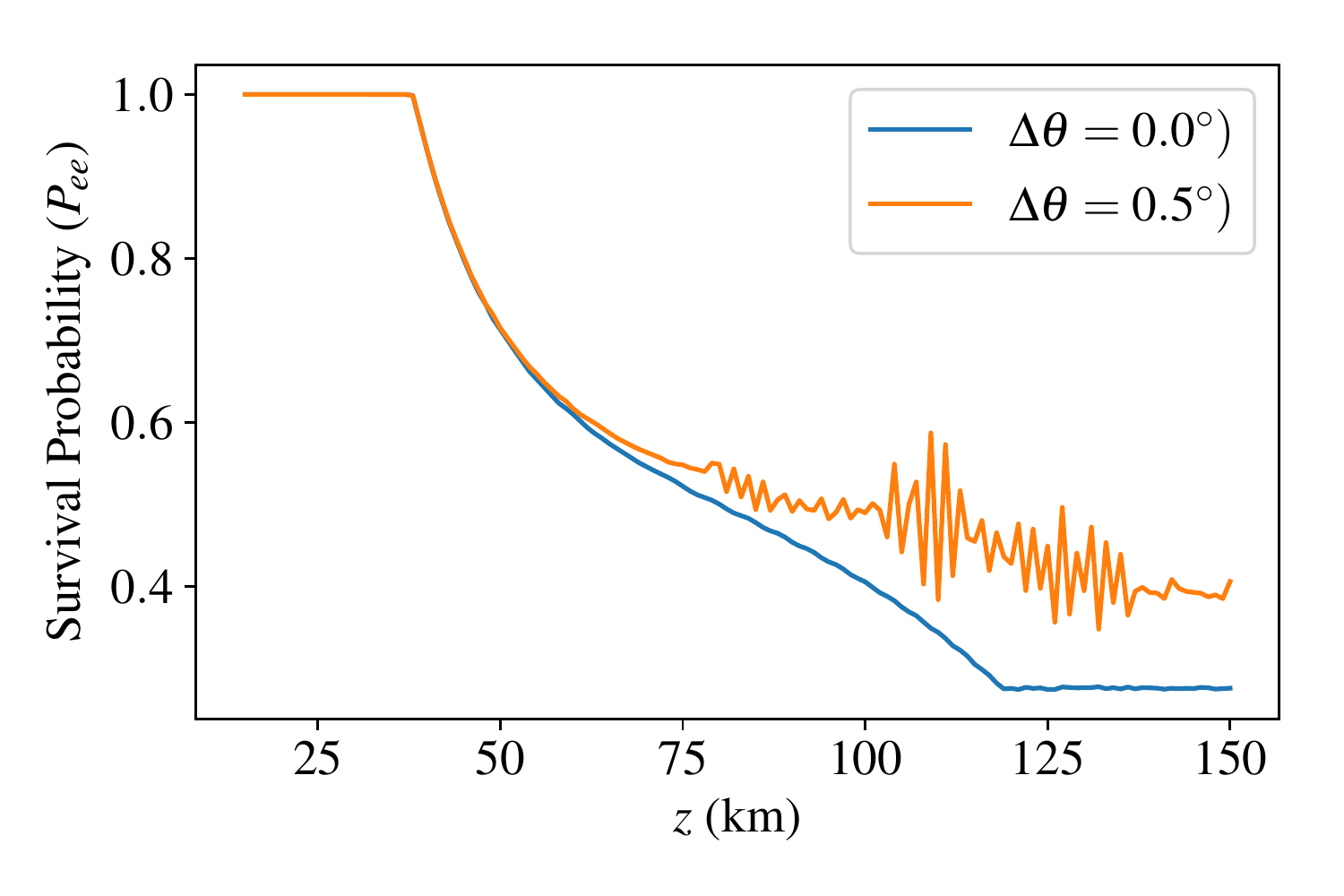}
\caption{The evolution of survival probability, $P_{ee}$, as a function of distance for electron neutrinos for normal hierarchy (left) and inverted hierarchy (right). The survival probability is plotted for the energy bin corresponding to the energy of 10 MeV. The two lines show the evolution for single-angle ($\Delta\theta=0.0^{\circ}$) and angle averaged survival probability for multi-angle calculation with $\Delta\theta=0.5^{\circ}$}
\label{surprob}
\end{figure}

\begin{figure}
\includegraphics[width=0.49\textwidth]{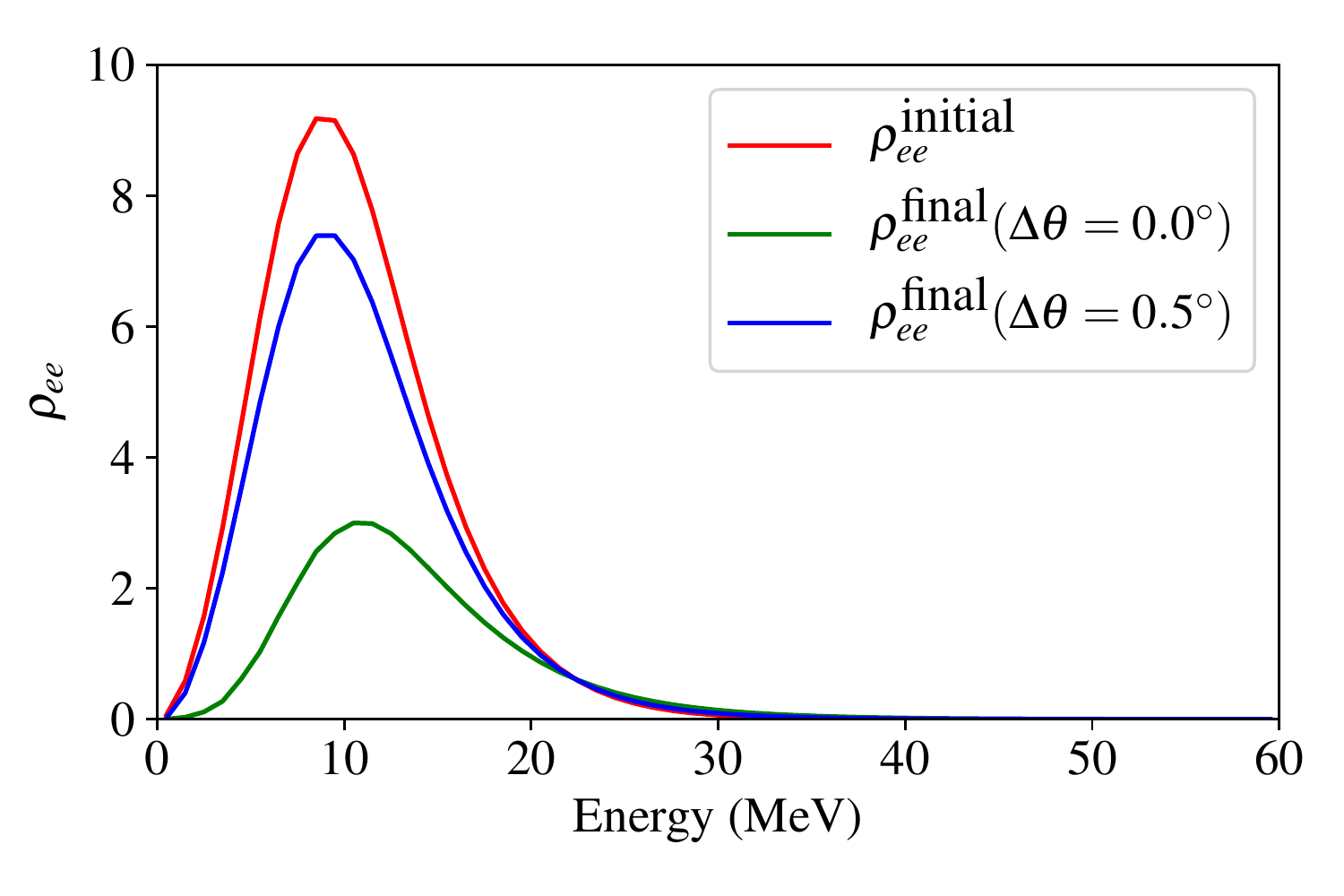}
\includegraphics[width=0.49\textwidth]{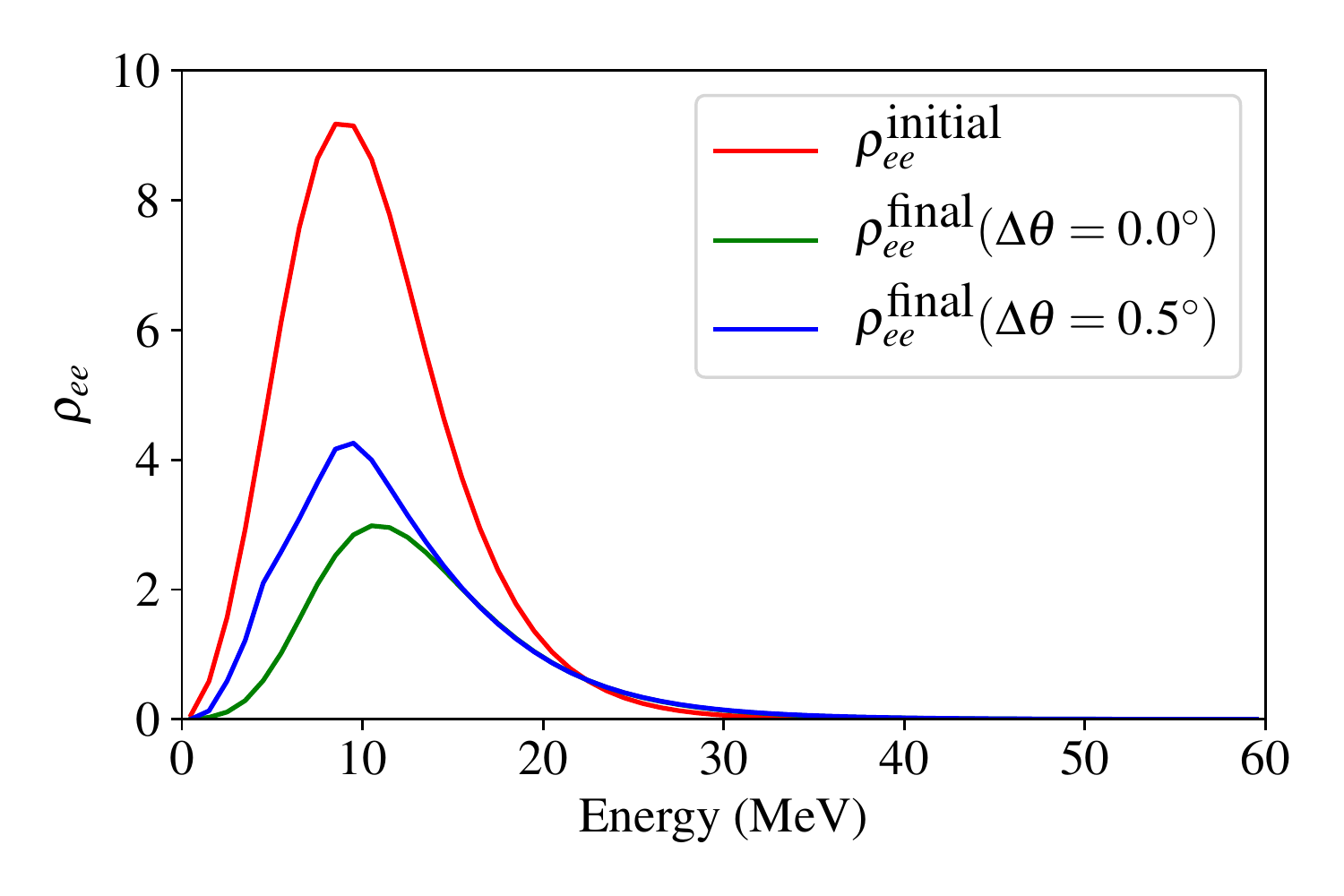}
\caption{Initial and final flux as a function of energy for normal(left) and inverted(right) hierarchy. The green line is the final flux in single-angle approximation and the blue line is the angle averaged final flux with an opening angle of $0.5^{\circ}$. The initial flux (red line) is plotted for reference. The plots show the evolution of the flux and have an arbitrary overall normalization.}
\label{finflux}
\end{figure}

In Fig.~\ref{surprob} we plot the survival probability for electron neutrinos for the energy bin corresponding to energy of 10 MeV. The relatively short lived matter-neutrino resonance in the case of multi-angle calculation with an opening angle of $0.5^{\circ}$ is clearly visible in the evolution of the angle averaged survival probability. Fig.~\ref{finflux} shows the comparison of the final electron neutrino fluxes for single-angle and angle averaged multi-angle calculations. The initial electron neutrino flux has been plotted in Fig.~\ref{finflux} for reference. It can be clearly seen in Fig.~\ref{surprob} and \ref{finflux} that the final angle averaged survival probability for neutrinos is larger in the case of multi-angle calculation compared to the single-angle approximation. As the opening angle is increased the matter-neutrino resonance becomes more and more short lived till it completely disappears.

The opening angles of interest in a physical system is determined by the ratio of the size of the accretion disk and the distance where neutrino oscillations can happen.
The interest in matter-neutrino resonance is driven by its possible influence on r-process nucleosynthesis, which occurs not too far from the accretion disk ($\lesssim 100$ km). The opening angle of around $2^{\circ}$ at which the matter-neutrino resonance disappears, is much smaller than the opening one would expect to encounter in any realistic calculation of neutrino flavor oscillations in the vicinity of accretion disks, assuming that the accretion disk cannot be less in 10 km in diameter.

\section{Discussion and conclusions}
\label{conclusion}
All the numerical calculations and qualitative discussions in the literature with regard to the phenomenon of matter-neutrino resonance are done in the context of single-angle approximation. In this paper we show that in the simplest of the models considered in the literature, the phenomenon of matter-neutrino resonance is an artifact of single-angle approximation and not a physical phenomenon.

It should be noted however, that we have done multi-angle calculations in simplest of the models of accretion disk. The conclusions drawn in this paper could be overturned by including more realistic physics. In this paper we have assumed that the electron number density is unchanged by neutrino oscillations both globally and locally. Strictly speaking this is not true, as neutrino oscillations can change the ratio of electron and anti-electron neutrinos which in turn affects neutron to proton ratio via $W$-boson exchange\footnote{The electron number density should be the same as proton number density to maintain the condition of vanishing electric charge for the accretion disk.}.

Also, in this paper we have assumed spatial homogeneity that is not broken by neutrino self-interactions. The validity of this assumption is in question in light of several studies on this topic~\cite{Duan:2014gfa,Abbar:2015mca,Mirizzi:2015fva,Chakraborty:2015tfa,Cirigliano:2017hmk}. Due to the complex geometry of an accretion disk it would be difficult to perform an analysis of matter-neutrino resonance for each and every Fourier mode of spatial imhomogeneity. It is possible, in principle, that the phenomenon of matter-neutrino resonance survives for small scale imhomogeneous modes even for large opening angles. In that case matter-neutrino resonance will only add to the richness of strange phenomena we expect to see due to neutrino self-interactions in astrophysical systems.

\acknowledgments

I would like to thank Huaiyu Duan for encouraging me to write this paper. Vincenzo Cirigliano and Mark Paris made several useful suggestions to improve the quality of the draft. I would like to thank Gail McLaughlin for interesting discussions on the topic of matter-neutrino resonance. This work was supported by LDRD program at Los Alamos National Laboratory. 


\bibliographystyle{JHEP}
\bibliography{mnr}
\end{document}